\documentclass[preprintnumbers,secnumarabic,amsmath,amssymb,nofootinbib,floatfix]{revtex4}
\usepackage{varioref,exscale,latexsym,amsmath,amssymb}
\usepackage{graphicx,epstopdf}
\usepackage{epsfig}
\usepackage{graphicx}
\usepackage{dcolumn}
\usepackage{bm}

\newcommand\slurp[1]{#1}
\newcommand\sslurp[2]{#1{#2}}
{\catcode`/=\active \expandafter}%
\slurp{\newcommand/}{/\penalty1000\hskip0pt\relax}
{\catcode`:=\active \expandafter}%
\slurp{\newcommand:}{:\penalty1000\hskip0pt\relax}

\newcommand\addspace{\ifcat\nextchar a\spacefactor999. \else.\fi}
{\catcode`\.=\active \expandafter}%
\slurp{\newcommand.}{\futurelet\nextchar\addspace}

\usepackage[unicode]{hyperref}
\ifx\href\undefined\def\href#1{}\fi
\ifx\texorpdfstring\undefined\def\texorpdfstring#1#2{#1}\fi

\newcommand\myslash{/} \newcommand\mycolon{:}
\newcommand\doi{{\catcode`/=\active \catcode`:=\active \expandafter}\sslurp\realdoi}
{\catcode`/=\active \catcode`:=\active \expandafter}%
\slurp{\newcommand\realdoi[1]{{\let/=\myslash \let:=\mycolon
                               \edef\raw{{http://dx.doi.org/#1}}\expandafter}%
                               \expandafter\href\raw{doi:#1}}}
\newcommand\eprint[2]{{\escapechar-1%
                       \edef\a{\expandafter\string\csname arXiv\endcsname}%
                       \edef\b{\expandafter\string\csname #1\endcsname}%
                       \edef\c{\expandafter\string\csname #2\endcsname}%
                       \edef\d{\noexpand\href{http://arXiv.org/abs/\c}}%
                       \ifx\a\b\expandafter\d\fi{\tt #1:#2}}}

\newcommand{\be}{\begin{equation}}
\newcommand{\ee}{\end{equation}}

\def\d{{\rm d}}
\def\OMIT#1{{}}

\newcommand{\eq}[1]{Eq.~\eqref{#1}}

\newcommand{\vc}[1]{\boldsymbol{#1}}

\def\SCETG{${\rm SCET}_{\rm G}\,$}

\begin{document}

\preprint{\vbox{\hbox{ACFI-T14-09}}}


\begin{flushright}
\end{flushright}

\title{{\bf Regge behavior in effective field theory}}

\medskip\
\author{ John F. Donoghue${}$}
\email[Email: ]{donoghue@physics.umass.edu}
\author{Basem Kamal El-Menoufi${}$}
\email[Email: ]{bmahmoud@physics.umass.edu}
\author{Grigory Ovanesyan${}$}
\email[Email: ]{ovanesyan@umass.edu}
\affiliation{Department of Physics,
University of Massachusetts\\
Amherst, MA  01003, USA\\
}

\begin{abstract}
We derive the Regge behavior for the forward scattering amplitude in scalar field theory using the method of regions. We find that the leading Regge behavior
to all orders can be obtained. Regge physics emerges from a kinematic region that involves the overlap of several modes, so that
a careful treatment of the overlap regions is important. The most consistent and efficient approach utilizes graphs
containing collinear, anti-collinear and Glauber modes, or modes of \SCETG.
\end{abstract}
\maketitle

\section{Introduction}
One indication that our effective field theory for high energy QCD is incomplete is that it presently does not reproduce Regge phenomena. This is dangerous because
Regge behavior can convert logarithms in a scattering amplitude into powers of the energy. In this paper we find Regge behavior in a related
effective field theory, and explore the modes that are needed to produce it.

The Soft Collinear Effective Theory (SCET)\cite{SCET},
is an effective field theory for QCD that is relevant for describing the dynamics of highly energetic quarks and gluons. In order to obtain
the Lagrangian of SCET one divides a single field into modes corresponding to distinct kinematic behavior and keeps leading terms consistent with the power counting.
Using these modes individually one can reconstruct the behavior of the Feynman diagrams of the full theory. Much of the insight into which modes to include has been obtained from original work by Collins, Soper and Sterman on all-order factorization theorem proofs \cite{Collins:1981ta} and from the method of regions \cite{MOR}. In this method one starts with the QCD Lagrangian and writes down an amplitude to a given order in the perturbation theory and expands it in one of the momentum regions that are identified using the pinch technique and power counting. Overviews with further references can be found in the book by Smirnov \cite{asymptotic} or the review by Jantzen \cite{jantzen}. Sometimes the individual modes are not fully kinematically distinct - there are overlap regions where more than one mode is active\cite{overlap}. These must
be carefully dealt with. It will turn out that the Regge physics lives in these overlap regions and involves a complicated interplay of regions and overlaps.
The most consistent and efficient way to describe it uses $\text{SCET}$ including Glauber modes \SCETG, as will be described below.

\begin{figure}[ht]
 \begin{center}
  \includegraphics[scale=0.8]{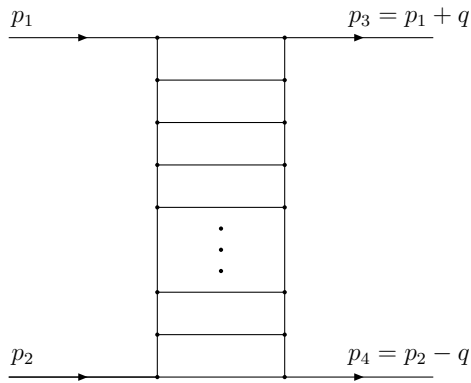}
 \end{center}
 \caption{\small{The ladder graphs }}
 \label{ladder}
\end{figure}

\begin{figure}[ht]
 \begin{center}
  \includegraphics[scale=0.8]{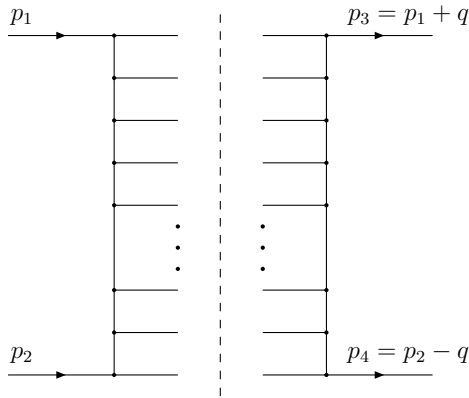}
 \end{center}
 \caption{\small{The cut ladder graphs }}
 \label{cut}
\end{figure}

The simplified model is that of a scalar field with a trilinear interaction, which can be considered a scalar model for QCD\footnote{For more on the relation of this model to Regge behavior in QCD, see the book by Forshaw and Ross\cite{Forshaw}. In this paper we will refer
to the full theory as QCD and the effective theory alternately as SCET without Glauber modes or \SCETG with them.}.
\begin{eqnarray}
\mathcal{L}=\frac{1}{2}\,\partial_{\mu}\phi\,\partial^{\mu}\phi-\frac{g}{3!}\phi^3.
\label{phi3}
\end{eqnarray}
The kinematic region where Regge behavior emerges is $s\to \infty, ~t$ fixed. In such theory the leading Regge behavior appears from summing an infinite set of ladder graphs, shown in Figure \ref{ladder} along with crossed ladder diagrams. The original calculation is due to Polkinghorne\cite{Polkinghorne, Eden}.
It is also very useful to know that we can reconstruct the Regge behavior of the ladder sum from consideration of the s-channel
discontinuities in the diagrams, where the relevant discontinuities are those where the cut lines are the rungs of
the ladder, as in Figure \ref{cut}. In the cut analysis, it is required that all the ladder rungs correspond to on-shell modes, so this fact needs to be
accommodated in the mode expansion.

\begin{figure}[t!]
\centering
\includegraphics[width=0.2\textwidth]{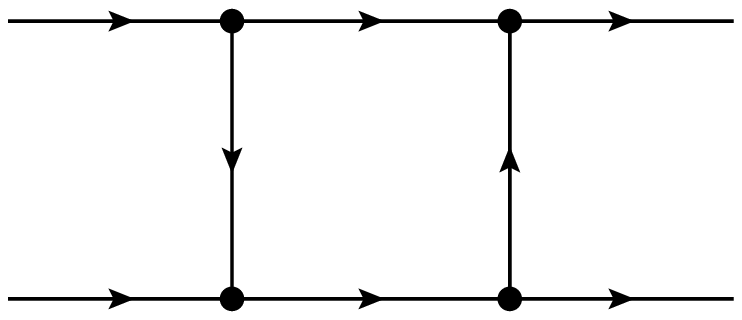}\quad\quad\quad\quad
\includegraphics[width=0.2\textwidth]{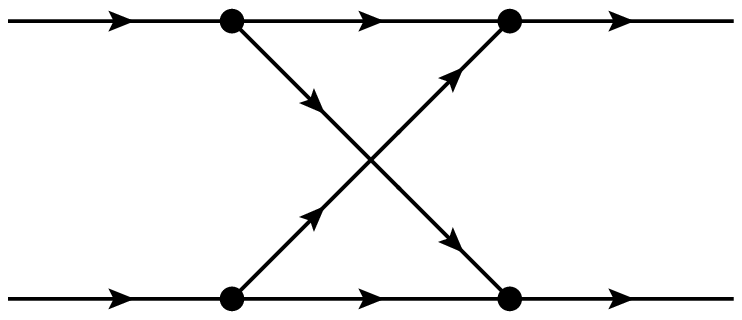}\quad\quad\quad\quad
\includegraphics[width=0.2\textwidth]{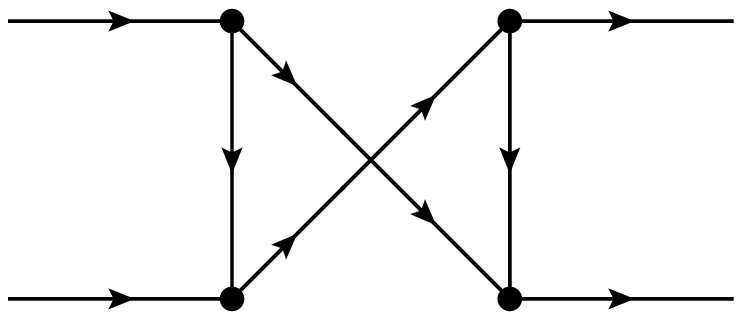}
\put(-380,45){$p_1$}
\put(-380,-5){$p_2$}
\put(-285,45){$p_3$}
\put(-285,-5){$p_4$}
\put(-360,20){$l$}
\put(-240,45){$p_1$}
\put(-240,-5){$p_2$}
\put(-145,45){$p_3$}
\put(-145,-5){$p_4$}
\put(-200,45){$p_1-l$}
\put(-100,45){$p_1$}
\put(-100,-5){$p_2$}
\put(-5,45){$p_3$}
\put(-5,-5){$p_4$}
\put(-80,20){$l$}
\caption{One-loop Feynman diagrams with box-like topology. We only show one internal momentum enough to clarify our conventions. The graphs represent the ($s,t$), ($u,t$) and ($s,u$) channels respectively. The last graph is suppressed by $t/s$ compared to first two.}
\label{box-topology}
\end{figure}

Our analysis will start with the mode expansion for the scalar box diagram, Figure \ref{box-topology}a. Along the way we will resolve a paradox that exists in
the usual method of regions treatment of the box. In \cite{asymptotic} Smirnov demonstrates how the box diagram can be reconstructed
by the use of collinear modes for the vertical legs of the ladder, although an extra ``analytic regularization'' in which
the propagators are modified is required. Indeed, we will also find this result with our regularization. However, when the legs are collinear modes, at least one of the horizontal rungs of the box must be a hard mode which is far off-shell. (We will review the terminology and kinematics in more detail below.) By unitarity, the off-shell mode should not be able to produce the
imaginary part of the box diagram. However, we will show that the imaginary part arises from an overlap region which the collinear mode shares with Glauber exchange. By removing the overlap, the box can be reformulated in a version of SCET including the Glauber mode, \SCETG, in which case the horizontal rung is in fact an on-shell (collinear) mode. The need to include Glauber modes in SCET has been shown by \cite{glauber} (see also \cite{jantzensmirnov}), they have been shown to be important in the context of jets in a medium\cite{medium}, and the relevance of these modes for Regge physics was first shown in \cite{wyler}.

The plan of this paper involves a brief overview of Regge behavior in Sec. 2, and of SCET kinematics in Sec. 3.
Then in Sec. 4 (along with Appendix A)
we provide a detailed treatment of the box diagram, paying particular attention to the overlap regions
between modes and demonstrating the importance of the Glauber mode. Sec. 5 treats the two loop ladder graph and
shows how to count the modes and match to the full theory. This is continued to higher orders in Sec. 6, 7.
A conclusion summarizes what has been accomplished. While this paper was being finalized, an important related work by
Fleming was released \cite{fleming}, and we also discuss the relation of our work to his in the
conclusion. Three appendices provide some relevant technical details.
\\


\section{Regge behavior in field theory}

For the purposes of this paper we will refer to Regge behavior as the dependence of the scattering amplitude on a power of the center of mass energy
\begin{equation}
{\cal M}_{\text{QCD}} \sim s^{\alpha(t)}
\end{equation}
in the limit $s \to \infty,~t$ fixed. The Regge exponent $\alpha (t)$ is dynamically generated through loop diagrams. At each order in perturbation
theory, the loops generate logs, but in this kinematic region the logs exponentiate into a power. In general one finds
\begin{equation}
{\cal M}_{\text{QCD}} \sim a_0 s^a \sum_{n=0}^\infty \frac{ \beta^n(t)}{n!}\ln^n s +... \to a_0 s^{a+\beta (t)} +.....
\label{sum}
\end{equation}
where we have allowed an extra possible overall factor of $s^a$ to the amplitude. (In our example $a=-1$.) It is this conversion of logs into powers that
makes the phenomenon important for phenomenology. In real QCD one sees a variety of Regge exponents depending on the quantum numbers, including the Pomeron with
$\alpha (0) \sim 1$.

Polkinghorne\cite{Polkinghorne} was the first to show how this behavior emerges in a field theory, using a massive scalar field with the $\phi^3$ interaction of Eq. \ref{phi3}. Although the ladder diagrams cannot be calculated completely, the leading
high energy behavior emerges from a corner of the Feynman parameter integration and this corner can be analyzed and summed. For example, the direct box diagram shown in Figure 3 after momentum integration becomes
\begin{equation}
i{\cal M}_{\text{QCD}}^{box}= i\frac{g^4}{16\pi^2}\int dx_1~dx_2~dy_1~dy_2 \frac{\delta (1-x_1-x_2-y_1-y_2)}{[x_1x_2 s + y_1y_2 t -m^2(1-(x_1+x_2)(y_1+y_2))]^2}
\end{equation}
where $x_1,~x_2$ are the Feynman parameters associated with the horizontal lines (rungs)  in the diagram and $y_1,~y_2$ are associated
with the vertical lines (legs). It is clear from this that the amplitude falls as $1/s^2$ at large $s$, except for the region of
integrations where $x_1$ and/or $x_2$ are close to zero. Polkinghorne noted that when only one of $x_{1,2}$ is small, the integrated amplitude
falls as $1/s$, but when both of the parameters are small there is an extra logarithmic factor of $\ln (-s)$. In this corner the
residual dependence on $x_{1,2}$ can be neglected and the result is
\begin{equation}
i{\cal M}_{\text{QCD}}^{box}= ig^2 \beta(t)
\frac{1}{-s}\ln (-s)
\end{equation}
with $s=s+i0$ and
\begin{equation}
\beta (t) = \frac{g^2}{16\pi^2}\int dy_1 dy_2 \frac{\delta (1-y_1-y_2)}{[m^2 -y_1y_2 t]} = \frac{g^2}{4\pi}\int \frac{d^2\vc{l}_{\perp}}{(2\pi)^2} \frac{1}{[\vc{l}_{\perp}^2+m^2][(\vc{l}_{\perp}+\vc{q}_{\perp})^2 +m^2]}
\label{beta}
\end{equation}
Note that the exponent depends on the transverse momenta only - the longitudinal components have been integrated out. The crossed box diagram is obtained
by the substitution $s\to u$, and since $s \gg -t, m^2$  we have $u\approx -s$. The sum of the box and crossed box then becomes
\begin{equation}
{\cal M}_{\text{QCD}}^{box~+~crossed}= -g^2 \beta(t)
\left[\frac{1}{s}\ln (-s)+ \frac{1}{-s}\ln (s)\right] = \frac{i\pi}{s} g^2\beta(t).
\end{equation}
This is the $n=0$ term in the Regge sum of Eq. \ref{sum}. In this case, we see that the result emerges entirely from the s-channel cut with both horizontal rungs being on-shell.

The rest of the ladder sum is done in the same way. The important region in the integration is the corner where all the Feynman parameters associated with the horizontal rungs becomes small. In this corner the correct $\ln^n(-s)$ behavior arises and the sum yields the Regge form with amplitude and exponent being
given by
\begin{equation}
a_0 = \frac{i\pi}{s} g^2\beta(t)  ~~~~~~~{\rm and }~~~~~~~~\alpha(t) =-1+\beta(t)  \ \ .
\end{equation}

In real QCD, the situation is somewhat more complicated, but follows the same kinematic rules. Within perturbative QCD,
this has been demonstrated by Balitsky, Fadin, Kuraev and Lipatov (BFKL) \cite{BFKL} and in related work\cite{collaborators}.

\section{Kinematics and notation}
We consider the binary scattering of particles with momenta $p_1$ and $p_2$, while the outgoing particles carry momenta $p_3$ and $p_4$. The momentum transfer is defined as $q=p_3-p_1$. We work in the center-of-mass frame and use light-cone coordinates requiring the introduction of two independent null vectors which read
\begin{align}
n^{\mu} = (1,0,0,1), \quad \bar{n}^{\mu}=(1,0,0,-1), \quad n \cdot \bar{n} = 2  \ \ .
\end{align}
Hence, four momenta are decomposed as follows
\begin{align}
p^{\mu} = p^+ \frac{\bar{n}^{\mu}}{2} + p^- \frac{n^{\mu}}{2} + \vc{p}_{\perp}, \quad p^+ \equiv p \cdot n, \quad p^- \equiv p \cdot \bar{n}, \quad \vc{p}_{\perp} \cdot n = \vc{p}_{\perp} \cdot \bar{n} = 0 \ \ .
\end{align}
For later convenience, we note the following identity
\begin{align}
\d^4 l = \frac{1}{2} \d l^+ \d l^- \d^2\vc{l}_{\perp} \ \ .
\end{align}

Regge physics is concerned with the kinematical limit
\begin{align}
s \rightarrow \infty, \quad -t,m^2 \ll s
\end{align}
where $s=(p_1+p_2)^2$ and $t=q^2$ are the usual Mandelstam variables. The small parameter required for employing the method of regions (SCET) then reads
\begin{align}
\lambda = \sqrt{\frac{-t}{s}}   \ \ .
\end{align}
All external particles are treated as massless and on-shell, in particular $p^2_i = 0$. Note that the scattering of two high energy on-shell particles, one in the $n$ direction and the other in the $\bar{n}$ always involves an exchange in the so-called Glauber region. This can be readily seen from the on-shell conditions
\begin{align}
p_3^2=0 &=(p_1+q)^2 = 0 + p_1^-q^+ +t   \nonumber \\
p_4^2=0 &=(p_2-q)^2 = 0 - p_2^+q^- +t   \ \ .
\label{onshell}
\end{align}
Because $p_1^-,p_2^+\sim \sqrt{s}$, this forces $q$ to scale as $q \sim \sqrt{s}(\lambda^2,\lambda^2,\lambda)$ in the $(+,-, \perp)$ directions. The Glauber region is characterized by having momentum dominantly in the transverse direction. The overall momentum transfer of Regge exchange is Glauber-like. In addition, one can include modes in the mode expansion which correspond to Glauber kinematics. Such modes are always off-shell, thus in the effective theory language they can be treated as an effective potential.

There is an array of possibilities in the choice of infrared regulators for our calculation. Among them is the analytic regulator used in \cite{asymptotic}, off-shellness of external momenta $p_i^2\ne 0$, or internal masses $m_i\ne 0$. If one uses off-shellness as a regulator with vanishing internal masses within the loop, one finds that the modes in the effective theory or method of regions are not regularized in four dimensions. Hence off-shellness by itself fails to regulate the infrared behavior of the theory and one needs to add dimensional regulator in order to regulate the infrared divergences.

In the remainder of this paper we regulate the infrared through the use of an internal mass for each internal line in any graph, keeping the external four-vectors on-shell with zero invariant mass. This fully controls the infrared region. For the leading high energy behavior the answer is the same if one uses massive external four-vectors with the
same $m^2$ as in the original Polkinghorne calculation.
\\

\section{One-loop box}

In this section we calculate the one-loop $\mathcal{O}(g^4)$ contribution to Regge physics of the binary scattering explained above. We start off by computing the appropriate graphs in the full theory and then repeat the calculation using the method of regions to isolate the modes responsible for Regge behavior. The graphs at the one-loop level which concern Regge physics are the ones with box-like topology shown in Figure \ref{box-topology}. In fact, the last graph has a suppressed leading behavior (by a power of $s$) compared to the first two and thus we neglect this graph all together. To fix the nomenclature, we refer to the first graph as the 'direct-box' and the second as 'crossed-box'.\\
\\
{\it{The box diagram in the full theory}}\\

The full Regge amplitude is simply obtained by summing the two graphs to find
\begin{align}
\nonumber
\mathcal{M}_{\text{QCD}}^{(1)} = (-i) g^4\frac{1}{2} \int \, \frac{\d^4l}{(2\pi)^4}\frac{1}{(l^2 - m^2 + i0) \, ((l+q)^2 - m^2 + i0)}&\left(\frac{1}{(l-p_1)^2 - m^2 + i0}+\frac{1}{(l+p_3)^2 - m^2 + i0}\right)\nonumber\\
\times  \,&\left(\frac{1}{(l+p_2)^2 - m^2 + i0}+\frac{1}{(l-p_4)^2 - m^2 + i0}\right)\, .
\end{align}
In the above expression we combined the graphs after symmetrizing each under the interchange $l \leftrightarrow -(l+q)$, and hence the factor of half. This does not prove useful for the full theory calculation but considerably simplifies the calculation in the method of regions. The intermediate steps of the computation are rather complicated and we move the details to Appendix A but the final result in the limit $s \gg -t,m^2$ takes the simple form
\begin{eqnarray}
\mathcal{M}_{\text{QCD}}^{(1)} =\frac{i \pi g^2 \beta(t)}{s},\label{eq:QCDoneloop}
\end{eqnarray}
where $\beta(t)$ is defined in \eq{beta} and explicitly reads
\begin{eqnarray}
\beta(t)=\frac{g^2}{8 \pi^2 \,(-t)\,\chi(t) }\ln\frac{\chi(t)+1}{\chi(t)-1}, \qquad \text{and }\qquad \chi(t)=\sqrt{1-\frac{4m^2}{t}}>1.\label{eq:betadef}
\end{eqnarray}
To arrive at this expression we have kept all the finite terms in the expansion $t/s, m^2/s$ and only dropped power-suppressed ones.
\\
\\
{\it{The box diagram in}} SCET {\it without Glauber modes}\\
\\
\begin{figure}[t!]
\centering
\includegraphics[width=0.25\textwidth]{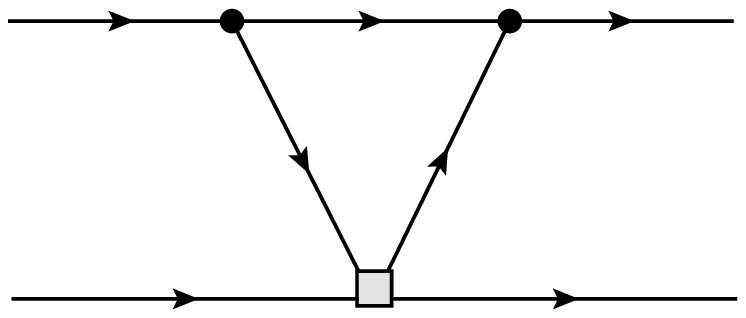}\quad\quad\quad\quad
\includegraphics[width=0.25\textwidth]{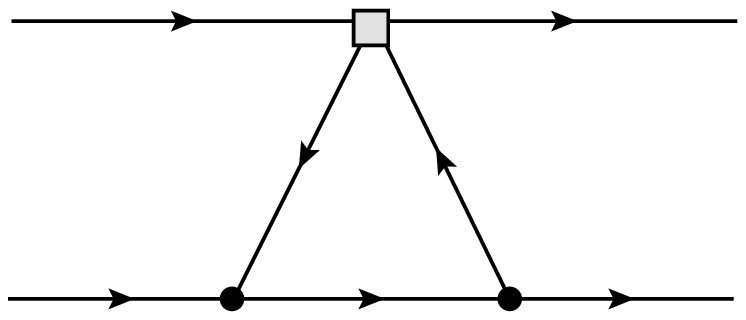} \quad\quad\quad\quad
\includegraphics[width=0.25\textwidth]{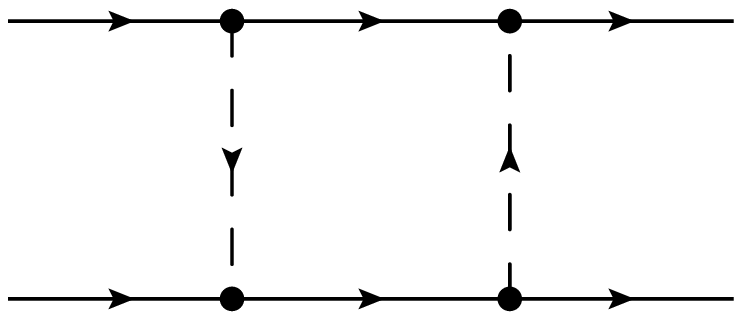}
\put(-455,60){$p_1$}
\put(-455,-10){$p_2$}
\put(-340,60){$p_3$}
\put(-340,-10){$p_4$}
\put(-292,60){$p_1$}
\put(-292,-10){$p_2$}
\put(-177,60){$p_3$}
\put(-177,-10){$p_4$}
\put(-122,60){$p_1$}
\put(-122,-10){$p_2$}
\put(-17,60){$p_3$}
\put(-17,-10){$p_4$}
\caption{'Direct-box' diagrams in $\text{SCET}_{\text{G}}$. The box represents an off-shell propagator and the dashed lines refer to Glauber modes. The momentum routing is identical to the box in Fig. \ref{box-topology}. In SCET, only the first two graphs appear.}
\end{figure}
Using the modes of SCET we get two leading graphs, when the loop momentum is either $n$-collinear or $\bar{n}$-collinear. An ultra-soft loop momentum $(\lambda^2,\lambda^2,\lambda^2)$ on the other hand is power suppressed because $m^2$ scales as $\lambda^2$. With some details in Appendix A we find that both $n$ and $\bar{n}$ collinear graphs shown in Figure 4 are equal
\begin{eqnarray}
\mathcal{M}_{n}^{(1)}=\mathcal{M}_{\bar{n}}^{(1)}=\mathcal{M}^{(1)}_{\text{QCD}}.
\end{eqnarray}
Hence, summing both contributions gives a result twice as big as the full theory. It turns out that the overlap contribution of these two modes is non-vanishing and must be taken into account in order to correctly reproduce the full theory result
\begin{align}
\mathcal{M}^{(1)}_{n/\bar{n}}=\mathcal{M}^{(1)}_{\text{QCD}}.
\end{align}
We derive a master formula in Appendix C that takes into account correct subtraction terms when combining $N$ momentum regions in the method of regions. Applying it for two modes, we reproduce the QCD result
\begin{align}
\mathcal{M}^{(1)}_{\text{SCET}}=\mathcal{M}^{(1)}_n+\mathcal{M}^{(1)}_{\bar{n}}-\mathcal{M}^{(1)}_{n/\bar{n}}=\mathcal{M}^{(1)}_{\text{QCD}}.
\end{align}
We have found that using the modes of SCET we recover the full QCD loop integral. This should not be surprising since for the box integral the pinch analysis leads to no Glauber pinch \cite{glauberQFT2,glauber}, i.e. modes of SCET are sufficient. However there is something strange with this result. Indeed the full answer for the QCD integral (defined by including the crossed box graph) is purely imaginary, see \eq{eq:QCDoneloop}. This imaginary part can be obtained from the discontinuity of the direct box integral. This discontinuity comes from on-shell intermediate states, however both collinear graphs $\mathcal{M}_n$ and $\mathcal{M}_{\bar{n}}$ have one propagator far off-shell. Thus, the imaginary part should come from sub-region of these two modes and is contained in a different kinematic region. In the book \cite{asymptotic} the one-loop box integral with the Regge kinematics was calculated using  analytic regulator. It was found that the box integral is recovered from two collinear graphs, similarly to our finding in this section with our regularization. Note, that with their regulator the overlap contribution vanishes and does not play role. Thus, the same comment that imaginary part of the box graph is coming from a different kinematic region and is outside of collinear graphs holds for the calculation in  \cite{asymptotic}. Below we repeat the one-loop calculation with our regulator by including all the modes of $\text{SCET}_{\text{G}}$ with their overlaps and this paradox is resolved. \\
\\
\\
\\
{\it The problem of the imaginary part}\\
\\
The imaginary part of the collinear amplitudes hints that we are missing insight into Regge physics. The imaginary part of the full theory should not be expected to come from collinear exchange. In simple words, this is because the collinear graphs indeed have one intermediate state off-shell. To elucidate this point, we directly employ Cutkosky rule to compute the imaginary part of the 'direct-box' graph. Hence,
\begin{align}\label{1loopim}
\text{Im}\, \mathcal{M}^{(1)}_{\text{QCD}} = \frac{g^4}{8\pi^2} \int \, d^4 l \frac{\delta_+ ((p_1-l)^2-m^2) \delta_+((l+p_2)^2-m^2)}{(l^2 - m^2)((l+q)^2-m^2)}
\end{align}
where $\delta_+(p^2-m^2) = \delta(p^2-m^2) \Theta(p^0)$. This integral is easily rewritten as
\begin{align}
\text{Im}\, \mathcal{M}^{(1)}_{\text{QCD}} = \frac{g^4}{16 \pi^2 \sqrt{s}} \int \, d^4 l \frac{\delta ((p_1-l)^2-m^2) \delta(l^0)}{\sqrt{s}(l^0-l_z)(\sqrt{s}(l^0-l_z)+ q^2 - 2 \vec{l} \cdot \vec{q})}
\end{align}
Notice that $q^0 = 0$ since we work in the center-of-mass frame. The integral over $l^0$ is readily done to absorb the second delta function and forces $l^0 = 0$. This is very interesting because this means that $l^{\pm} = \mp l_z$ which shows that the mode responsible for generating the leading Regge behavior ought to have longitudinal components of equal scaling; a condition clearly violated by collinear modes. We continue the calculation by performing the $l_z$ integral where a quadratic form appears in the argument of the delta function with the following roots
\begin{align}
l_z^{\pm} = \frac{\sqrt{s}}{2} \left(1 \pm \sqrt{1 - \frac{4 \Delta}{s}}\right), \quad \Delta = \vc{l}^2_{\perp} + m^2
\end{align}
The '$+/-$' refers to large/small root respectively. Clearly, the transverse integral has to be constrained since $4\Delta \leq s$ for $l^{\pm}_z$ to be real. The large root yields a result proportional to $1/s^2$, and hence is power-suppressed. On the other hand, the small root can be Taylor-expanded
\begin{align}
l_z^- = \frac{\Delta}{\sqrt{s}}\left(1 + \frac{\Delta}{s} + ... \right)
\end{align}
The first term in the expansion yields the leading result in the Regge limit, and hence $l_z^- \approx \Delta / \sqrt{s}$. This is the second piece of information we need to pin-down the Regge mode; it has an excess in the transverse direction identical to the momentum transfer. We conclude that Glauber scaling $l \sim \sqrt{s}(\lambda^2,\lambda^2,\lambda)$ is genuinely resposible for Regge behavior. Finally, the result agrees with the full calculation
\begin{align}
\text{Im}\, \mathcal{M}^{(1)}_{\text{QCD}} = \frac{\pi g^2 \beta(t)}{s}
\end{align}
\\
{\it{The box diagram in \SCETG}}\\
\\
In Soft Collinear Effective Theory with Glauber modes an additional graph appears where the loop momentum is that of Glauber scaling $l(\lambda^2,\lambda^2,\lambda)$. With details provided in Appendix A, the box integral in this momentum region is equal to
\begin{align}
\mathcal{M}_G^{(1)}  = \mathcal{M}^{(1)}_{\text{QCD}}  \ \ .
\end{align}
When adding the Glauber as an independent mode, overlaps with collinear modes must be taken into account, we calculate these in Appendix A. The result is
\begin{align}
\mathcal{M}^{(1)}_{n/G}  = \mathcal{M}^{(1)}_{\bar{n}/G} =\mathcal{M}^{(1)}_{n/\bar{n}/G}= \mathcal{M}_G^{(1)}=\mathcal{M}^{(1)}_{\text{QCD}} \ \ .
\end{align}
Our calculation indicates that the imaginary part of the full theory is coming precisely from the Glauber region (at the one-loop order). In other words the matching in \SCETG after all zero-bin (overlap) subtractions gives the same result as just the Glauber mode
\begin{align}
\mathcal{M}^{(1)}_{\text{\SCETG}}  = \mathcal{M}_n^{(1)}+\mathcal{M}_{\bar{n}}^{(1)}+\mathcal{M}_G^{\text{box}}-\mathcal{M}_{n/\bar{n}}^{(1)}-\mathcal{M}_{n/G}^{(1)}-\mathcal{M}_{\bar{n}/G}^{(1)}+\mathcal{M}_{n/\bar{n}/G}^{(1)}=  \mathcal{M}^{(1)}_{G} =\mathcal{M}^{(1)}_{\text{QCD}}.
\end{align}
As is well-known from factorization proofs of the Drell-Yan process, the Glauber region is not pinched for the box topology \cite{glauberQFT2}, (see also \cite{glauber}). Thus, no surprise that we get the same answer in $\text{SCET}_{\text{G}}$ as in $\text{SCET}$. When one uses the modes of SCET, the collinear integrals each contain at least one of the intermediate propagators off-shell, thus the imaginary part\footnote{Note that entire one-loop expression of the QCD amplitude is imaginary, see \eq{eq:QCDoneloop}.} ought to come from a subregion inside them. Our calculation above precisely shows that the correct momentum region for the full box integral in the Regge kinematics (when direct and crossed box are added) is the {\it{Glauber}} region. What we mean by this is that the QCD box integral with crossed box added at one loop can be reproduced by a single Glauber mode, that does not violate unitarity. As one adds the collinear graphs and Glauber one together, the interpretation can be made that the true collinear mode obtained from naive collinear mode after zero-bin subtraction becomes purely real (as it should be due to unitarity) and cancels out between box and crossed box. Thus we resolve the paradox of imaginary part coming from collinear graphs. We will see below in this paper that this generalizes straightforwardly to higher orders in perturbation theory.
\\
\\
{\it The imaginary part via Cutkosky rule}
\\
\\
For completeness and as a prelude to the next section, we directly use Cutkosky rule to recalculate the imaginary part albeit taking the loop momentum in the Glauber region. The computation is very simple and the correct result is obtained effortlessly. Expanding the integrand of (\ref{1loopim}) and explicitly employing light-cone coordinates
\begin{align}
\text{Im}\, \mathcal{M}^{(1)}_{G} = \frac{g^4}{16\pi^2} \int \, \d^2 \vc{l}_{\perp} \, \d l^+ \d l^- \frac{\delta(-\vc{l}^2_{\perp}-\sqrt{s}l^+-m^2) \delta(-\vc{l}^2_{\perp}+\sqrt{s}l^--m^2)}{(\vc{l}^2_{\perp} + m^2)((\vc{l}_{\perp}+\vc{q}_{\perp})^2+m^2)}  \ \ .
\end{align}
The step functions are automatically satisfied. The longitudinal momenta integrals are used trivially to absorb the delta functions and we find
\begin{align}
\text{Im}\, \mathcal{M}^{(1)}_{G} = \frac{ \pi g^2 \beta(t)}{s}   \ \ .
\end{align}
As previously mentioned, the kinematic exercise of Eq. \ref{onshell} combined with the Cutkosky rule for the on-shell intermediate states of the box diagram require that the exchanged
mode be in the Glauber region\footnote{We note that the analytic regulator used in \cite{asymptotic} sets the Glauber region to zero. It is then not consistent with the Cutkosky rule and we view this as a disadvantage of this regulator.}.
\\
\\
\section{Two-loop ladder}
We saw at one-loop that the underlying physics behind Regge behavior is the imaginary part of the 'direct-box' graph caused by the s-channel discontinuity. This picture is valid at any loop order as confirmed by Polkinghorne analysis \cite{Polkinghorne}. Hence, we confine the subsequent discussion to only the imaginary part of higher loop graphs that contribute to Regge physics, i.e. ladder graphs.\\
\\
\begin{figure}[t!]
\centering
\includegraphics[width=0.3\textwidth]{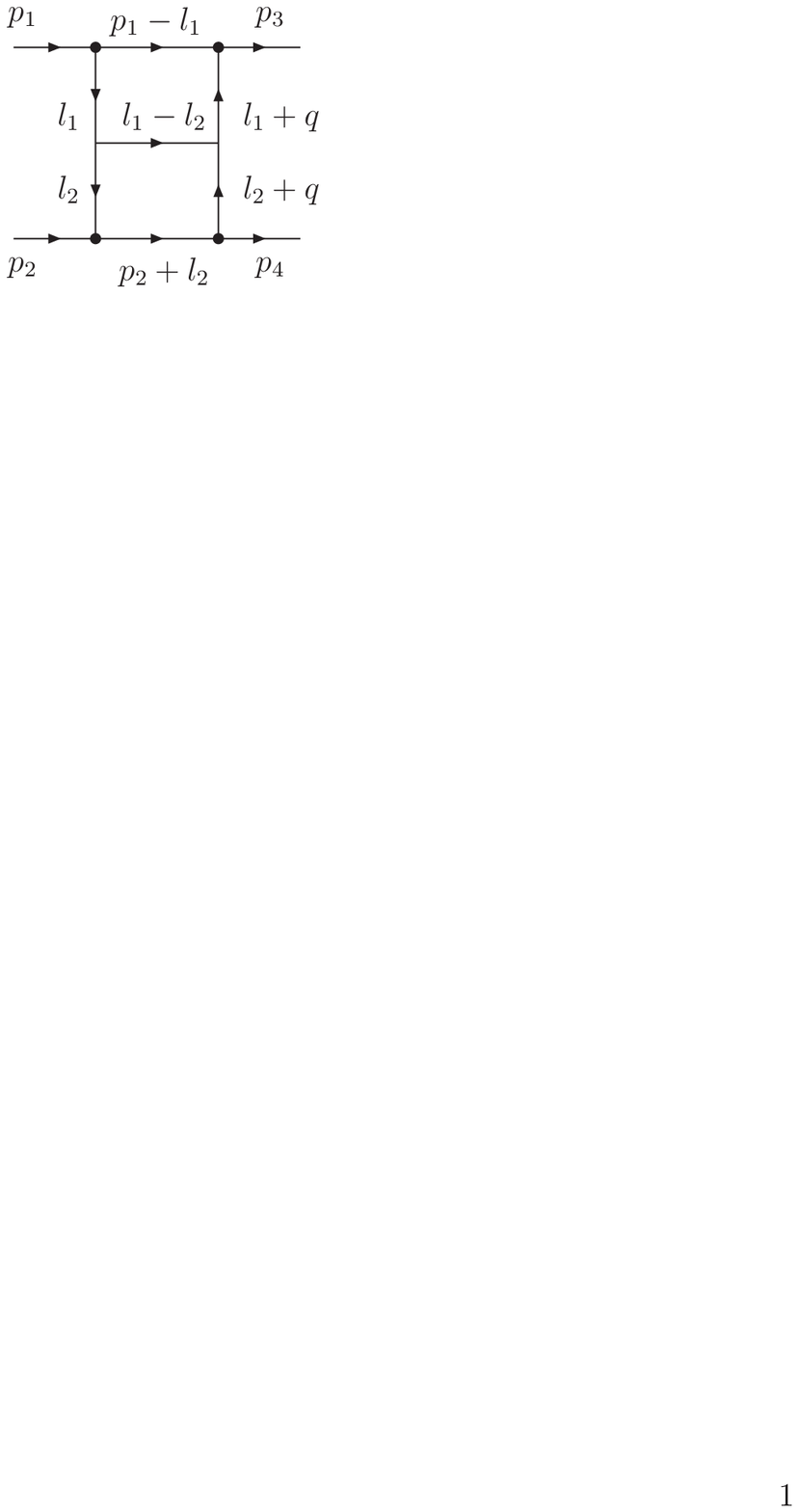}
\caption{Two-loop direct graph in full theory (QCD).}
\end{figure}
{\it{The double box ladder in the full theory}}\\
\\
The imaginary part of the two-loop ladder graph shown in Figure 5 is given by
\begin{align}\label{2loop}
\text{Im} \, \mathcal{M}^{(2)}_{\text{QCD}} = \frac{g^6}{64 \pi^5} \int \, &d^4l_1 d^4l_2 \, \frac{\delta_+[(p_1 - l_1)^2-m^2]  \delta_+[(l_1-l_2)^2-m^2] \delta_+[(p_2 + l_2)^2-m^2]}{(l^2_1-m^2) (l^2_2-m^2) ((l_1+q)^2-m^2) ((l_2+q)^2-m^2)},
\end{align}
where $\delta_+$ is defined as before. This integral has extensively been studied before with the result
\begin{eqnarray}
\text{Im}\,\mathcal{M}^{(2)}_{\text{QCD}} \approx\frac{\pi g^2 \beta^2(t)}{s} \ln  s\, .
\end{eqnarray}
For example, see the derivation in \cite{Forshaw,wyler}, where it is shown that the leading region is the so-called strongly ordered one. The formula above contains the leading logarithm as $s\rightarrow \infty$. There are finite terms beyond this logarithm that are same order in the $t/s$ expansion and thus are expected to be captured by the method of regions or the mode expansion. For this reason we derived such terms in Appendix B, but for simplicity we have set $t=0$. This leads to a finite answer in four dimensions because our massive theory is infrared safe. The result is
\begin{align}\label{2loopanswer}
\text{Im} \, \mathcal{M}^{(2)}_{\text{QCD}}(t=0) = \frac{g^6}{256 \pi^5} \int \, &d^2\vc{l}_{1\perp}\d^2\vc{l}_{2\perp}\frac{1}{s\,(\Delta_1\Delta_2)^2}\left[\ln\frac{s}{\Delta_{12}}-2\right]\,\theta\left(s-\left(\sqrt{\Delta_1}+\sqrt{\Delta_2}+\sqrt{\Delta_{12}}\right)^2\right).
\end{align}
In the equation above we have made the following definitions in terms of transverse momenta
\begin{eqnarray}
\Delta_1=\vc{l}_{1\perp}^2+m^2, \qquad \Delta_2=\vc{l}_{2\perp}^2+m^2,\qquad \Delta_{12}=(\vc{l}_{1\perp}-\vc{l}_{2\perp})^2+m^2.
\end{eqnarray}
\begin{figure}[t!]
\centering
\includegraphics[width=0.3\textwidth]{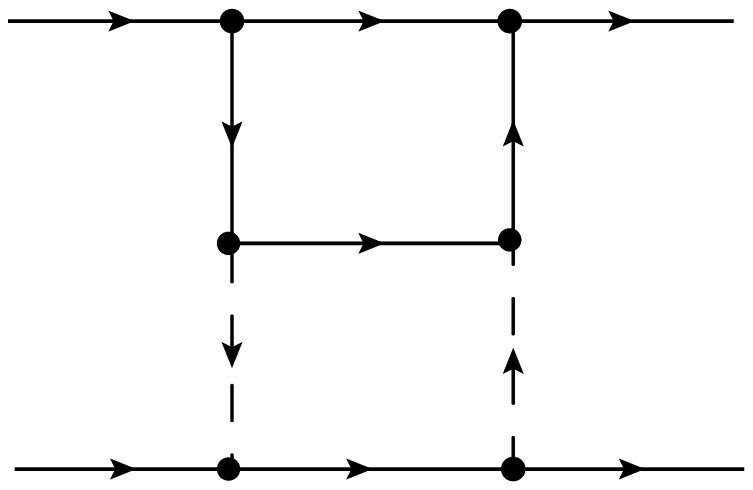} \qquad\qquad \includegraphics[width=0.3\textwidth]{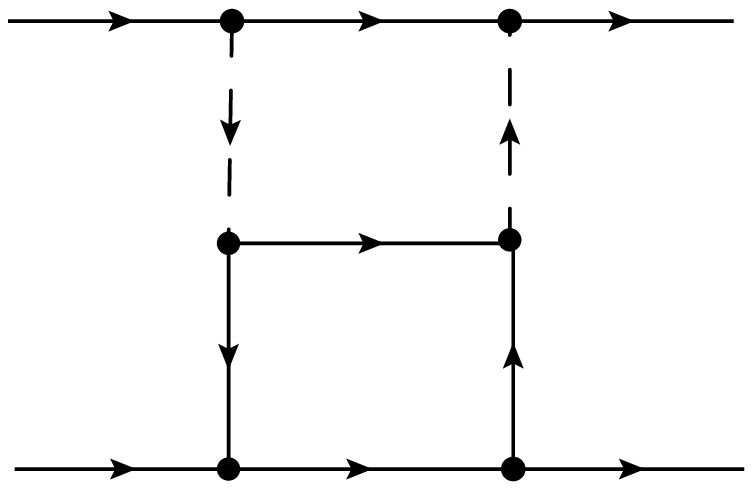}
\put(-345,102){$p_1$}
\put(-345,10){$p_2$}
\put(-210,102){$p_3$}
\put(-210,10){$p_4$}
\put(-147,102){$p_1$}
\put(-147,10){$p_2$}
\put(-12,102){$p_3$}
\put(-12,10){$p_4$}
\caption{Two-loop graphs in $\text{SCET}_{\text{G}}$. From left to right these amplitudes are $\mathcal{M}_{nG}, \,\mathcal{M}_{G\bar{n}}$. All other two-loop graphs in method of regions; for example $\mathcal{M}_{\bar{n} n}$, are power suppressed or lead to intermediate off-shell propagators, such as $\mathcal{M}_{nn}$.}
\end{figure}
\\
\\
{\it{Matching in \SCETG}}
\\
\\
So far we have learned from the one-loop calculation that the correct momentum region to understand the Regge behavior is when all intermediate states are on-shell. The only way to do this at two loops is one of the three possibilities $(l_1, l_2)$ is $(n,G),(G,\bar{n}),(G,G)$. Power-counting the mode integrals shows that $(G,G)$ must be suppressed, due to higher power of the momentum-space volume factor $d^4l_G\sim\lambda^6$ as opposed to  $d^4l_\text{coll}\sim\lambda^4$. The fact that there are two leading modes (shown in Figure 6) means that their overlap must be taken into account. The first of our modes equals to:
\begin{align}\label{2loopGnbar}
\text{Im} \, \mathcal{M}^{(2)}_{nG} = \frac{g^6}{64 \pi^5} \int \, &d^4l_1 d^4l_2 \, \frac{\delta_+[(-l_1^+)(\sqrt{s}-l_1^-)-\Delta_1] \delta_+[(l_1^+-l_2^+)l_1^--\Delta_{12}]  \delta_+[\sqrt{s}\,l_2^--\Delta_2] \,\theta(\sqrt{s}+l_2^+)}{(l_1^+l_1^--\Delta_1)(-\Delta_2) ((l_1^++q^+)\,l_1^--\Delta_{1q}) (-\Delta_{2q})},
\end{align}
where the theta function in each of the $\delta_+$ that appear above is the {\it{expanded}} expression in the QCD integral. We note that with such strict manifest power counting the limits of integration on the $l_1^-$ after the $l_1^+,l_2^+,l_2^-$ integrals are performed using the delta functions are not regulated and lead to divergent expression. To overcome this shortcoming we restore the {\it{unexpanded}} theta function in the $\delta_+[(p_2+l_2)^2-m^2]$ present in QCD expression. We made this explicit by inserting $\theta(\sqrt{s}+l_2^+)$ in the \eq{2loopGnbar}. We also made the following definitions
\begin{eqnarray}
\Delta_{1q}=(\vc{l}_{1\perp}+\vc{q}_{\perp})^2+m^2,\qquad \Delta_{2q}=(\vc{l}_{2\perp}+\vc{q}_{\perp})^2+m^2.
\end{eqnarray}
With details explained in Appendix B we derive expression for this mode for arbitrary values of $t$:
\begin{align}\label{2loopGnbar}
\text{Im} \, \mathcal{M}^{(2)}_{nG} = \frac{g^6}{256 \pi^5} \int \, &d^2\vc{l}_{1\perp}\d^2\vc{l}_{2\perp}\frac{1}{s\,\Delta_1\Delta_2\Delta_{1q}\Delta_{2q}}\left[\ln\frac{s}{\Delta_{12}}+\frac{1}{2}\ln\frac{\Delta_{1q}}{\Delta_{1}}-\frac{\arctan U}{U}\right]\,\theta\left(s-(\sqrt{\Delta_{12}}+\sqrt{\Delta_{1}})^2\right),
\end{align}
where
\begin{eqnarray}
U=\sqrt{\frac{4\Delta_1\Delta_{1q}}{(t+\Delta_{1}+\Delta_{1q})^2}-1}.
\end{eqnarray}
In the entire domain of integration over $\vc{l}_{1\perp}, \vc{l}_{2\perp}$ the value of $U>0$ and the answer is well-behaved. By the same argument as above it is clear that in the leading limit in the power expansion as $s\rightarrow \infty$ the theta function in the \eq{2loopGnbar} can be ignored. Taking the leading logarithmic limit of \eq{2loopGnbar} we get:
\begin{align}
\text{Im}\, \mathcal{M}^{(2)}_{nG} \approx \frac{\pi g^2 \beta^2(t)}{s} \ln  s.
\end{align}

Calculation of the second mode $(G,\bar{n})$ proceeds analogously and leads to the  identical result:
\begin{eqnarray}
\text{Im}\, \mathcal{M}^{(2)}_{G\bar{n}}=\text{Im}\, \mathcal{M}^{(2)}_{nG}.
\end{eqnarray}
The fact that both modes yield identical result can also be seen from the following change of variables in the loop integrals:
\begin{eqnarray}
l^{\pm}_{1}\leftrightarrow -l^{\mp}_2,\qquad \vc{l}_{1\perp}\leftrightarrow \vc{l}_{2\perp}.
\end{eqnarray}
This change of variables transforms integrand of $(G,\bar{n})$ mode with the one of the $(n,G)$ mode. Note that both modes that we considered reproduce exactly the leading Regge behavior of QCD, so if one adds them together the result is that of twice of QCD. We should remember from our one loop computation that overlaps need to be included, and thus we proceed to calculate the $(nG,G\bar{n})$ overlap.\\
\\
\\
{\it{The Regge mode}}
\\
\\
Here, we show that the overlap is the generator of Regge physics. Expanding the integrand of (\ref{2loop}) subsequently with scaling of the modes $nG$ and $G\bar{n}$, all propagators become transverse and factor out of the longitudinal integration
\begin{align}\label{2loopoverlap}
\text{Im} \, \mathcal{M}^{(2)}_{nG/G\bar{n}} = \frac{g^6}{64 \pi^5} \int \, &d^4l_1 d^4l_2 \, \frac{\delta_+[(-l_1^+)\sqrt{s}-\Delta_1] \delta_+[(-l_2^+)l_1^--\Delta_{12}]  \delta_+[\sqrt{s}\,l_2^--\Delta_2] \,\theta(\sqrt{s}-l_1^-)\,\theta(\sqrt{s}+l_2^+)}{(-\Delta_1)(-\Delta_2)(-\Delta_{1q})(-\Delta_{2q})}.
\end{align}
Note that the theta functions in $\delta_+$ are also {\it{expanded}} in this momentum region and similarly to the $nG$ mode considered above we inserted additional theta functions present in the full QCD expression that help regulate integrals. Using the delta functions yields:
\begin{align}
\bar{l}^+_1 = -\frac{\Delta_1}{\sqrt{s}}, \quad \bar{l}^-_2 =\frac{\Delta_2}{\sqrt{s}}, \quad \bar{l}^+_2 =  -\frac{\Delta_{12}}{l_2^+}.
\end{align}
As a result the overlap integral is equal to
\begin{align}\label{2loopoverlapsecondversion}
\text{Im} \, \mathcal{M}^{(2)}_{nG/G\bar{n}} = \frac{g^6}{256 \pi^5}\frac{1}{s} \int \, &d^2\vc{l}_1 d^2\vc{l}_2 \, \frac{1}{\Delta_1\Delta_2\Delta_{1q}\Delta_{2q}}\,I_{l_1^-}\,,
\end{align}
where
\begin{align}
\nonumber
I_{l_1^-} &= \int_{{\Delta_{12}}/{\sqrt{s}}}^{\sqrt{s}} \, \d l_1^- \, \frac{1}{l_1^-}\\
&= \ln\left(\frac{s}{\Delta_{12}}\right)  \, .
\end{align}
Thus, the overlap correctly captures the sub-region $l_1^- \ll \sqrt{s}$ and thus the final result in the Regge limit reads
\begin{align}
\text{Im} \, \mathcal{M}^{(2)}_{nG/G\bar{n}} = \frac{\pi g^2 \beta^2(t)}{s} \ln s\, .
\end{align}
This is same answer as in Regge limit of QCD.
\\
\\
{\it{Combining the modes of $\text{SCET}_\text{G}$}}
\\
\\
Combining all the modes in the effective theory under consideration, we get:
\begin{eqnarray}
&&\text{Im} \, \mathcal{M}^{(2)}_{\text{SCET}_{\text{G}}}=\text{Im} \, \left(\mathcal{M}^{(2)}_{nG}+\mathcal{M}^{(2)}_{G\bar{n}}-\mathcal{M}^{(2)}_{nG/G\bar{n}}\right)\nonumber\\
&& \qquad\qquad\,= \frac{g^6}{256 \pi^5} \int \, d^2\vc{l}_{1\perp}\d^2\vc{l}_{2\perp}\frac{1}{s\,\Delta_1\Delta_2\Delta_{1q}\Delta_{2q}}\left[\ln\frac{s}{\Delta_{12}}+\ln\frac{\Delta_{1q}}{\Delta_{1}}-2\frac{\arctan U}{U}\right]\,.
\end{eqnarray}
This combined answer reduced to the case $t=0$ reproduces the QCD result exactly
\begin{eqnarray}
&&\text{Im} \, \mathcal{M}^{(2)}_{\text{SCET}_{\text{G}}}(t=0)= \frac{g^6}{256 \pi^5} \int \, d^2\vc{l}_{1\perp}\d^2\vc{l}_{2\perp}\frac{1}{s\,\left(\Delta_1\Delta_2\right)^2}\left[\ln\frac{s}{\Delta_{12}}-2\right]\nonumber\\
&&\qquad\qquad\qquad\qquad\,\,\,=\text{Im} \, \mathcal{M}^{(2)}_{\text{QCD}}(t=0).
\end{eqnarray}
It also reproduces the leading Regge behavior of the QCD integral for arbitrary $t$. The fact that the leading Regge behavior is present in both modes and in the overlap as well, simply means that the true region from which the leading logarithm is coming is the overlap region. This result must be intimately connected to the strong ordering in the leading Regge limit. Indeed, strong ordering is a very special region in the momentum space with hierarchy of energies, and we argue that our observation that the leading mode is the overlap has the same roots. In the remaining sections we provide  arguments why this conclusions persists to higher orders.
\section{Three-loop ladder}
In this section, we demonstrate that the full overlap between all 'on-shell' modes immediately yields the leading Regge behavior similar to the two-loop case. At three loops we find three leading modes; $(n,n,G)$ $(n,G,\bar{n})$, $(G,\bar{n},\bar{n})$. The imaginray part of the three-loop ladder is obtained via Cutkosky rule and the expression is similar to (\ref{2loopoverlap}). The three-fold overlap between the leading modes forces all the propagators to become transverse and factor out as before. A close look at the expansion of the arguments of the delta functions for this multiple overlap momentum region leads to:
\begin{eqnarray}
&&\text{Im}\, \mathcal{M}^{(3)}_{nnG/nG\bar{n}/G\bar{n}\bar{n}}\nonumber\\
&&=\frac{g^8}{512\pi^8}\int \d^4 l_1\,\d^4 l_2\,\d^4 l_3\frac{\delta_+(-\sqrt{s}\,l_1^+-\Delta_1)\delta_+(-l_1^-l_2^+-\Delta_{12})\delta_+(-l_2^- l_3^+-\Delta_{23})\delta_+(\sqrt{s}\,l_3^--\Delta_3)}{\Delta_{1}\Delta_{2}\Delta_{3}\Delta_{1q}\Delta_{2q}\Delta_{3q}}\nonumber\\
&&=\frac{g^8}{512\pi^8}\frac{1}{2^3}\frac{1}{s}\int \d^2\vc{l}_{1\perp}\,\d^2\vc{l}_{2\perp}\,\d^2\vc{l}_{3\perp}\frac{1}{\Delta_{1}\Delta_{2}\Delta_{3}\Delta_{1q}\Delta_{2q}\Delta_{3q}}\,I_{l_1^-l_2^-}.
\end{eqnarray}
Notice the nice feature that the appearance of $\Delta_{i}$ terms inside the $\delta$ functions follows from the consistent power counting of the multi-overlap region. We use the delta function to integrate over all plus components in addition to $l_3^-$, and the remaining non-trivial integrals read\footnote{The prescription adopted to get these limits of integration, relies on {\it{unexpanded}} theta functions adopted from full QCD expression.}
\begin{align}
I_{l_1^-l_2^-}&=\int_{\Delta/\sqrt{s}}^{\sqrt{s}}\, \frac{\d l_2^-}{l_2^-} \int_{l_2^-}^{\sqrt{s}}\, \frac{\d l_1^-}{l_1^-}\\
&= \frac{1}{2} \ln^2\left(\frac{s}{\Delta}\right),
\end{align}
where $\Delta$ is a function of transverse momenta, but in the leading logarithmic approximation the answer does not depend on it. Finally, we get keeping only leading in the Regge limit term:
\begin{eqnarray}
\text{Im}\, \mathcal{M}^{(3)}_{nnG/nG\bar{n}/G\bar{n}\bar{n}}=\frac{\pi\,g^2\beta(t)^3}{s}\frac{\ln^2 s}{2}.
\end{eqnarray}
This matches the QCD result.

\section{Generalization to all orders}
From our explicit calculations at one and two loop orders it is easy to guess the answer for higher orders. We first guess the answer and then prove it further below. We expect that the true momentum region for the Regge kinematics at $N$ loop order is $N$ leading graphs which are subset of \SCETG graphs with on-shell intermediate states. These are the graphs with a number of $n-$collinear gluons in the loops, a single Glauber gluon, and after it a number of $\bar{n}-$collinear momentum in the loops
\begin{eqnarray}
\mathcal{M}^{(N)}_{n\dots nG}, \,\mathcal{M}^{(N)}_{n\dots nG\bar{n}\dots\bar{n}}, \dots\,\mathcal{M}^{(N)}_{G\bar{n}\dots\bar{n}}.
\end{eqnarray}
Each of this amplitudes includes by our definition both the direct box and the crossed box. Each of these amplitudes reproduces the leading Regge behavior as $s\rightarrow \infty$ and also any overlap of any subset of these amplitudes reproduces the Regge behavior. Thus once one combines all the modes, the answer is identical to only including a single mode which is the overlap of all $N$ momentum regions.

In order to prove the above statements we use the strong ordering derivation to show that arbitrary graph in the method of regions gives an identical result as single loop integral in QCD. Consider for example $\mathcal{M}^{(N)}_{G\bar{n}\dots\bar{n}}$ graph. The loop momenta $l_{i}$ where $i=1\dots N$ scale as $(l_{i}^{+},l_{i}^{-},\vc{l}_{i\perp})\sim (\lambda^2,\lambda^2,\lambda),\,(1,\lambda^2,\lambda), \,\dots (1,\lambda^2,\lambda) $. Thus the plus momentum satisfies $l_1^+\sim \lambda^2\ll l_2^+\sim\dots \sim l_{k}^+ \sim l_{k+1}^+\sim\dots \sim l_{N}^+\sim 1$ and $l_1^-\sim l_2^-\sim\dots \sim l_{k}^- \sim l_{k+1}^-\sim\dots \sim l_{N}^-\sim \lambda^2$. Clearly the strong ordering region is a subregion of this region, since for the strong ordered region we have\footnote{Note, that in this expression all the $``+"$ components are negative and all the $``-"$ components are positive. This is imposed by the theta functions in the expression for the QCD cut graph.}
\begin{eqnarray}
&& |l_1^+|\ll |l_2^+|\dots \ll |l_{k}^+| \ll |l_{k+1}^+|\ll\dots \ll |l_{N}^+|,\nonumber\\
 &&l_1^-\gg l_2^-\dots \gg l_{k}^- \gg l_{k+1}^-\gg\dots \gg l_{N}^-.
\end{eqnarray}
Thus repeating the usual strong ordering region derivation we would presumably get the same answer as in the full theory if we started to work on the graph $\mathcal{M}^{(N)}_{G\bar{n}\dots\bar{n}}$. Similarly we can show that every other relevant graph is identical to one another, since they all contain the strong ordering region as their sub-region.

An analogous statement holds for any of the loop integrals involving Glauber gluons. These subset of graphs are the only ones out of entire set that allow on-shell intermediate states. Our observation that the multi-overlap of these regions plays an important role has a simple interpretation. At $N-$loop order the single isolated momentum region that gives the leading Regge behavior is the multi-overlap of all on-shell modes $n\dots nG/n\dots n G\bar{n}\dots\bar{n}/\dots/G\bar{n}\dots\bar{n}$. It is easy to verify by a straightforward calculation similar to what we did at three-loop order
\begin{eqnarray}
\text{Im}\, \mathcal{M}^{(N)}_{n\dots nG/n\dots n G\bar{n}\dots\bar{n}/\dots/G\bar{n}\dots\bar{n}}=\frac{\pi\,g^2\beta(t)}{s}\frac{(\beta(t)\ln s)^{N-1}}{(N-1)!},
\end{eqnarray}
which reproduces the QCD Regge limit. In this section we showed that all leading modes have strong ordering momentum region as their sub-region, thus including only the multiple overlap of all these modes is sufficient and no surprise leads to the correct answer.
\section{Conclusions}

We have shown how one obtains Regge physics using the mode expansion of SCET. In the effective field
theory, the key contributions come from overlap regions which must be carefully treated. The simplest and
most consistent approach includes the Glauber modes of the effective field theory \SCETG.

In the scalar theory that we discuss, the one loop contribution that starts the Regge ladder sum
comes from the imaginary part of the box diagram. The box diagram can be reproduced in an effective
theory which includes only the hard and collinear modes. However this comes at the cost of seemingly
violating the unitarity property of field theory in that the imaginary part of the amplitude arises
from a hard intermediate state which the effective theory says is far off-shell. This result tell us
that in fact the contribution comes from an overlap region with an on-shell mode. By including the exchange of Glauber
modes in the description, we can again recover the full box diagram via the mode expansion. In this case,
after accounting for the overlap regions, the imaginary part of the amplitude is properly obtained from
t-channel Glauber exchange with s-channel on-shell collinear modes.

At higher order the deconstruction of the various overlap regions continues, with a final result
that is simple to state. Collinear modes provide many of the legs in the ladder sum, and all of
the s-channel on-shell states. However, at any given loop order, a Glauber mode is responsible
for the connection between the collinear $n$ and $\bar{n}$ modes. We have explicitly demonstrated
this at two loops, and provided an argument that this continues for all higher loops.

As we were writing up our project, a related paper by Fleming showed how one can obtain the well-known BFKL equation from rapidity RGE in the effective theory\cite{fleming}. In Fleming's calculation the forward scattering matrix element which falls in the Glauber region is calculated to one-loop order and rapidity divergent couterterm is determined. As a result the BFKL equation and the Regge behavior emerge from SCET with Glauber gluons. In our paper we also showed how Regge behavior emerges from SCET with Glauber gluons, however we looked at it from a fixed order point of view by summing a series of graphs order-by order in the perturbation theory. In doing so we paid a special attention to the momentum region where the Regge behavior arises from and we concluded that the Glauber mode plays an important role to connect the $n-$collinear sector with $\bar{n}-$ collinear sector with keeping intermediate propagators on-shell. The combined Regge mode we find is the mutual overlap of all such graphs at a given order and reproduces the Regge behavior.  It should be noted that Fleming uses real QCD and works in the framework of $\text{SCET}_{\text{II}}$ with Glaubers. In our work we only considered the modes of  $\text{SCET}_{\text{I}}$ with Glauber gluons added and used a toy scalar field theory.

 Neither this work nor \cite{fleming} (nor the early work of \cite{wyler}) is the final word on this subject. It is a common practice to apply SCET to resum Sudakov logarithms in high energy processes to very high orders in the logarithmic accuracy. The effective theory language is particularly efficient operationally. We hope that the results of this paper will guide the construction of a consistent EFT formulation of QCD that explains Regge behavior and allows the resummation of Regge logs similarly to existing techniques for Sudakov resummation. We need a technology that starts from the effective Lagrangian which allows a theorist to provide complete descriptions of processes including the usual SCET calculations but in addition including Regge contributions
when appropriate. The good news is that we can now see that Regge physics can be compatible with the effective
field theory. However, we do not yet have the complete technology to include such effects in realistic
calculation in a transparent and consistent fashion. From this work it follows that $\text{SCET}_{\text{G}}$ is an obvious candidate for such an effective field theory. We will pursue such approach in future work.

\vspace{9mm}
{\Large\bf Acknowledgements\/}\\\\
We thank Iain Stewart, Martin Beneke, Thomas Becher and Bernd Jantzen for useful conversations. JFD thanks the Institut des Hautes {\'E}tudes Scientifiques, the Niels Bohr International Academy and the University of Zurich for hospitality in early phases of this work. The work of JFD and BKM was supported in part by NSF grants PHY-0855119 and PHY-1205986, and that of GO by DOE grant DE-SC0011095.

\vspace{9mm}
\appendix
\section{The box diagram at one-loop order}
\subsection{Full theory (QCD)}
In this appendix, we list the calculational details of the box diagram in the full theory. We employ conventional Feynmann parametrization and integrate over the loop momentum to find within our regularization scheme
\begin{align}
\mathcal{M}^{box}_{\text{QCD}} = \frac{g^4}{16 \pi^2} \int_0^1 \, dx_1 dx_2 dy_1 dy_2 \, \frac{\delta(1-x_1-x_2-y_1-y_2)}{[x_1 x_2 s + y_1 y_2 t -m^2]^2}
\end{align}
where it is understood that
\begin{align}
s=s+i0, \quad m^2 = m^2 - i0\,  \ \ .
\end{align}
The result can be expressed in terms of four basic integrals
\begin{align}
\mathcal{M}^{box}_{\text{QCD}} = \frac{g^4}{16\pi^2} [I_1(s,t,m^2)+I_2(s,t,m^2)+I_1(t,s,m^2)+I_2(t,s,m^2)]
\end{align}
where
\begin{align}
\nonumber
I_1(a,b,m^2) = \int_0^1 dy \, \frac{\ln(a(1-y)-by)-\ln(-by)}{y(1-y)ab-m^2a-y^2b^2}, \quad I_2(a,b,m^2) = \int_0^1 dy \frac{\ln(m^2-y(1-y)b)-\ln(m^2)}{y(1-y)ab-m^2a-y^2b^2} \ \ .
\end{align}
We list the results of the basic integrals in the Regge limit
\begin{align}
\nonumber
I_1(s,t,m^2) = \frac{1}{st\alpha} &\bigg[\ln\left(\frac{s}{-t}\right)\ln\left(\frac{r_0^+}{r_0^-}\right)^2\bigg] \\\nonumber
I_2(s,t,m^2) = \frac{1}{st\alpha} &\bigg[\ln\left(\frac{-t}{m^2}\right)\ln\left(\frac{r_0^+}{r_0^-}\right)^2+\ln^2 l^+ - \ln^2(-l^-) - 2 \ln(\alpha)\ln\left(\frac{-l^-}{l^+}\right)+2\text{Li}_2 \left(\frac{-l^-}{\alpha}\right)-2\text{Li}_2 \left(\frac{l^+}{\alpha}\right)\bigg]\\\nonumber
I_1(t,s,m^2) = \frac{1}{st\alpha} &\bigg[\ln\left(\frac{s}{-t}\right)\ln\left(\frac{r_0^-}{r_0^+}\right)^2+\text{Li}_2 \left(\frac{1}{r_0^-}\right)-\text{Li}_2 \left(\frac{1}{r_0^+}\right) - \text{Li}_2 \left(\frac{s}{-t r_0^-}\right) + \text{Li}_2 \left(\frac{s}{-t r_0^+}\right)\\\nonumber
&+\text{Li}_2 \left(\frac{s}{t r_0^+}\right)-\text{Li}_2 \left(\frac{s}{t r_0^-}\right)\bigg]\\\nonumber
I_2(t,s,m^2) = \frac{1}{st\alpha} &\bigg[-\ln\left(\frac{-s}{m^2}\right)\ln\left(\frac{r_0^-}{r_0^+}\right)+\text{Li}_2 \left(\frac{m^2}{m^2 - tr_0^+}\right)-\text{Li}_2 \left(\frac{m^2}{m^2 - tr_0^-}\right)+\text{Li}_2 \left(\frac{s}{tr_0^- - m^2}\right)-\text{Li}_2 \left(\frac{s}{tr_0^+ - m^2}\right)\\\nonumber
&-i\pi \left(\ln\left(\frac{-t r_0^+}{s}\right)-\ln\left(\frac{t r_0^-}{s}\right)+\ln\left(\frac{s}{m^2 - t r_0^-}\right)-\ln\left(\frac{s}{m^2 -t r_0^+}\right)\right)\\
&+\ln\left(\frac{m^2}{s}\right)\left(\ln\left(\frac{tr_0^+}{tr_0^+ - m^2}\right)-\ln\left(\frac{tr_0^-}{tr_0^- - m^2}\right)\right)\bigg]  \ \ ,
\end{align}
where
\begin{align}
r_0^{\pm} = \frac{1}{2}(1\pm \chi(t)), \quad \chi(t) = \sqrt{1-\frac{4m^2}{t}} \ \ .
\end{align}
We note the important simplification
\begin{align}
\ln\left(\frac{tr_0^{\pm}}{tr_0^{\pm} - m^2}\right)=\ln\left(\frac{1}{r_0^{\pm}}\right) \ \ .
\end{align}
The crossed-box amplitude reads
\begin{align}
\mathcal{M}^{cross}_{\text{QCD}} = \frac{g^4}{16\pi^2} \int_0^1 \, dx_1 dx_2 dy_1 dy_2 \, \frac{\delta(1-x_1-x_2-y_1-y_2)}{[x_1 x_2 u + y_1 y_2 t -m^2]^2}  \ \ .
\end{align}
In the Regge limit, $u \approx -s$ and hence
\begin{align}
\mathcal{M}^{cross}_{\text{QCD}} = \frac{g^4}{16\pi^2} [I_1(-s,t,m^2)+I_2(-s,t,m^2)+I_1(t,-s,m^2)+I_2(t,-s,m^2)] \ \ .
\end{align}
Upon summing both amplitudes, all terms cancel except the dilogarithms whose arguments approache infinity in the Regge limit. Those must be expanded and handled carefully, with the final result being
\begin{align}
\mathcal{M}_{\text{QCD}} = \frac{i g^4}{8 \pi st\chi(t)} \ln\left(\frac{\chi(t)-1}{\chi(t)+1}\right)  \ \ .
\end{align}
\subsection{Modes in the EFT}
Expanding the one-loop box integral in the $n$-collinear region we get
\begin{align}
\nonumber
\mathcal{M}_{n}^{(1)} = (-i)g^4\frac{1}{2} \int \,\frac{\d^4l}{(2\pi)^4} &\frac{1}{(l^2 - m^2 + i0) \, (l^-(l+q)^+ -(\vc{l}_{\perp}+\vc{q}_{\perp})^2 - m^2 + i0)}\\
&\times\left(\frac{1}{(l^- -\sqrt{s})l^+ -\vc{l}_{\perp}^2- m^2 + i0}+\frac{1}{(l^- + \sqrt{s})(l+q)^+ -(\vc{l}_{\perp}+\vc{q}_{\perp})^2- m^2 + i0}\right)\nonumber\\
&\times  \left(\frac{1}{\sqrt{s} l^- + i0}+\frac{1}{-\sqrt{s} l^- + i0}\right)\, .
\end{align}
This integral nicely collapses upon using the following identity
\begin{align}
\frac{1}{x+i\epsilon} -\frac{1}{x-i\epsilon} = -2i\pi\,\delta(x) \ \ .
\end{align}
The $\bar{n}$-collinear region yields an identical result, and we find
\begin{eqnarray}
\mathcal{M}_{n}^{(1)}=\mathcal{M}_{\bar{n}}^{(1)}=\mathcal{M}^{(1)}_{\text{QCD}} \ \ .
\end{eqnarray}
The overlap contribution of these two modes is non-vanishing and must be taken into account in order to correctly reproduce the full theory result
\begin{align}
\mathcal{M}^{(1)}_{n/\bar{n}}=(-i)g^4\frac{1}{2} \int \, &\frac{\d^4{l}}{(2\pi)^4}\frac{1}{(l^2 - m^2 + i0) \, (l^+l^- -(\vc{l}_{\perp}+\vc{q}_{\perp})^2 - m^2 + i0)}\nonumber\\
&\Big(\frac{1}{-\sqrt{s}l^- + i0}+\frac{1}{\sqrt{s}l^- + i0}\Big) \times  \Big(\frac{1}{\sqrt{s} l^+ + i0}+\frac{1}{-\sqrt{s} l^+ + i0}\Big)\, \ \ ,
\end{align}
with the final result
\begin{align}
\mathcal{M}^{(1)}_{n/\bar{n}}=\mathcal{M}^{(1)}_{\text{QCD}} \ \ .
\end{align}

In Soft Collinear Effective Theory with Glauber modes an additional graph appears where the loop momentum is that of Glauber scaling $l(\lambda^2,\lambda^2,\lambda)$. The box integral in that momentum region reads
\begin{align}
\nonumber
\mathcal{M}_G^{(1)}  = (-i)g^4\frac{1}{2} \int \, &\frac{\d^4l}{(\vc{l}_{\perp}^2 + m^2 - i0) \, ((\vc{l}_{\perp} + \vc{q}_{\perp})^2 + m^2 - i0)}\\\nonumber &\times \Big(\frac{1}{(-\vc{l}_{\perp}^2 - \sqrt{s} \, l \cdot \bar{n} - m^2 + i0)} + \frac{1}{(-\vc{l}_{\perp}^2 + \sqrt{s} \, l \cdot \bar{n} - 2 \vc{l}_{\perp} \cdot \vc{q}_{\perp} - m^2 + i0)}\Big)\\
&\times \Big(\frac{1}{(-\vc{l}^2_{\perp} + \sqrt{s} \, l \cdot n - m^2 + i0)} + \frac{1}{(-\vc{l}_{\perp}^2 - \sqrt{s} \, l \cdot n - 2 \vc{l}_{\perp} \cdot \vc{q}_{\perp} - m^2 + i0)}\Big)\, .
\end{align}
The integral is elementary and yields
\begin{align}
\mathcal{M}_G^{(1)}  = \mathcal{M}^{(1)}_{\text{QCD}} \ \ .
\end{align}
When adding the Glauber as an independent mode, overlaps with collinear modes must be taken into account
\begin{align}
&\mathcal{M}^{(1)}_{n/G}= (-i)g^4\frac{1}{2} \int \, \frac{\d^4l}{(\vc{l}_{\perp}^2 + m^2 - i0) \, ((\vc{l}_{\perp} + \vc{q}_{\perp})^2 + m^2 - i0)}\nonumber  \\
&\qquad\qquad\qquad\qquad\times \left(\frac{1}{- \sqrt{s} \, l \cdot \bar{n}+ i0} + \frac{1}{ \sqrt{s} \, l \cdot \bar{n} + i0}\right)\nonumber\\
&\qquad\qquad\qquad\qquad\times \left(\frac{1}{(-\vc{l}^2_{\perp} + \sqrt{s} \, l \cdot n - m^2 + i0)} + \frac{1}{(-\vc{l}_{\perp}^2 - \sqrt{s} \, l \cdot n - 2 \vc{l}_{\perp} \cdot \vc{q}_{\perp} - m^2 + i0)}\right)\, ,\\
&\mathcal{M}^{(1)}_{\bar{n}/G}= (-i)g^4\frac{1}{2} \int \, \frac{\d^4l}{(\vc{l}_{\perp}^2 + m^2 - i0) \, ((\vc{l}_{\perp} + \vc{q}_{\perp})^2 + m^2 - i0)}\nonumber  \\
&\qquad\qquad\qquad\qquad\times \Big(\frac{1}{(-\vc{l}_{\perp}^2 - \sqrt{s} \, l \cdot \bar{n} - m^2 + i0)} + \frac{1}{(-\vc{l}_{\perp}^2 + \sqrt{s} \, l \cdot \bar{n} - 2 \vc{l}_{\perp} \cdot \vc{q}_{\perp} - m^2 + i0)}\Big)\nonumber\\
&\qquad\qquad\qquad\qquad\times \Big(\frac{1}{\sqrt{s} \, l \cdot n  + i0} + \frac{1}{ - \sqrt{s} \, l \cdot n + i0}\Big)\, ,\\
&\mathcal{M}^{(1)}_{n/\bar{n}/G}= (-i)g^4\frac{1}{2} \int \, \frac{\d^4l}{(\vc{l}_{\perp}^2 + m^2 - i0) \, ((\vc{l}_{\perp} + \vc{q}_{\perp})^2 + m^2 - i0)}\nonumber  \\
&\qquad\qquad\qquad\qquad\times \left(\frac{1}{- \sqrt{s} \, l \cdot \bar{n}  + i0} + \frac{1}{+ \sqrt{s} \, l \cdot \bar{n} + i0}\right)\nonumber\\
&\qquad\qquad\qquad\qquad\times \left(\frac{1}{+ \sqrt{s} \, l \cdot n + i0} + \frac{1}{( - \sqrt{s} \, l \cdot n  + i0}\right)\, .
\end{align}
The result of each is identical to the Glauber integral
\begin{align}
\mathcal{M}^{(1)}_{n/G}  = \mathcal{M}^{(1)}_{\bar{n}/G} =\mathcal{M}^{(1)}_{n/\bar{n}/G}= \mathcal{M}_G^{(1)}=\mathcal{M}^{(1)}_{\text{QCD}}\, .
\end{align}
\section{The ladder diagram at two-loop order}
\subsection{Full theory (QCD)}
Using Cutkosky's rules the imaginary part of the two-loop ladder graph in the full theory (QCD) can be written as:
\begin{eqnarray}\label{2loopAppendix}
\text{Im} \, \mathcal{M}^{(2)}_{\text{QCD}} = \frac{g^6}{64 \pi^5} \int \, d^4l_1 d^4l_2 \, \frac{\delta_+[(p_1 - l_1)^2-m^2]  \delta_+[(l_1-l_2)^2-m^2] \delta_+[(p_2 + l_2)^2-m^2]}{(l^2_1-m^2) (l^2_2-m^2) ((l_1+q)^2-m^2) ((l_2+q)^2-m^2)},
\end{eqnarray}
where $\delta_+(p^2-m^2)=\theta(p^0)\,\delta(p^2-m^2)$. One can work out the result of performing integration over $l_1^+,l_2^+,l_2^-$ integrals using the three delta functions. Working out the delta and theta functions for $t=0$ leads to the following:
\begin{eqnarray}
\text{Im} \, \mathcal{M}^{(2)}_{\text{QCD}}= \frac{g^6}{256 \pi^5}\frac{1}{s}\int\d^2\vc{l}_{1\perp}\d^2\vc{l}_{2\perp}\frac{\theta\left(s-\left(\sqrt{\Delta_1}+\sqrt{\Delta_2}+\sqrt{\Delta_{12}}\right)^2\right)}{\left(\Delta_1\Delta_2\right)^2}\,I_{l_1^-},
\end{eqnarray}
 where
 \begin{eqnarray}
 I_{l_1^-}=\frac{1}{s}\int_{y_{\text{min}}}^{y_{\text{max}}}\d l_1^-\frac{\sqrt{s}-l_1^-}{l_1^-}\frac{1}{|x_1-x_2|}\sum_{i=1}^{2}(\sqrt{s}+x_i)^2.\label{eq:integralI1master}
 \end{eqnarray}
In the equation above the limits of integration are dictated by the theta functions in the cut diagram and $y_{\text{min}}$ and $y_{\text{max}}$ are the smallest and biggest of the roots of the quadratic equation:
 \begin{eqnarray}
\sqrt{s}\,y\,(\sqrt{s}-y)=\Delta_1\,y+\left(\sqrt{\Delta_2}+\sqrt{\Delta_{12}}\right)^2\left(\sqrt{s}-y\right)\, .
\end{eqnarray}
In the equation \eq{eq:integralI1master} $x_1, x_2$ are the two roots of the quadratic equation (in $x$):
\begin{eqnarray}
\left(l_1^- - \frac{\Delta_2}{\sqrt{s}+x}\right)\left(x+\frac{\Delta_1}{\sqrt{s}-l_1^-}\right)+\Delta_{12}=0\, .
\end{eqnarray}
Integrand in the equation \eq{eq:integralI1master} is equal to fourth order polynomial in the $l_1^-$ divided by a square root of a fourth order polynomial in the denominator and divided by $\left(l_1^-\right)^2$. The integrand simplifies if one keeps all the roots of the numerator and denominator (in $l_1^-$) to the leading order in $s\rightarrow \infty$. Then one gets:
 \begin{eqnarray}
I_{l_1^-} \approx \frac{1}{\sqrt{s}}\int_{\alpha_2}^{\alpha_3}\frac{\d l_1^-}{\left(l_1^-\right)^2}\frac{\left(l_1^--\beta_1\right)\left(l_1^--\beta_2\right)\left(\alpha_3-l_1^-\right)}{\sqrt{\left(l_1^--\alpha_1\right)\left(l_1^--\alpha_2\right)}},\label{eq:integral1expanded}
 \end{eqnarray}
where
\begin{eqnarray}
\alpha_1=\frac{\left(\sqrt{\Delta_{12}}-\sqrt{\Delta_{2}}\right)^2}{\sqrt{s}},\qquad \alpha_2=\frac{\left(\sqrt{\Delta_{12}}+\sqrt{\Delta_{2}}\right)^2}{\sqrt{s}},\qquad \alpha_3=\sqrt{s},\qquad \beta_{1,2}=\frac{\Delta_{12}\pm\sqrt{\Delta_2}\sqrt{2\Delta_{12}-\Delta_2}}{\sqrt{s}}.\label{eq:alpharoots}
\end{eqnarray}
Performing the integral above and keeping leading $s\rightarrow \infty$ term gives
\begin{eqnarray}
I_{l_1^-}\approx \ln\frac{s}{\Delta_{12}}-2.\label{eq:i1answer}\ \ .
\end{eqnarray}
The last integral can be done exactly and expanded. It can also be noticed that since the $\Delta_1$ dependence dropped out from all roots in \eq{eq:alpharoots}, and because the original expression was symmetric in $\Delta_1$ and $\Delta_2$ one can could have guessed that the answer is independent on $\Delta_2$ to the leading order as $s\rightarrow \infty$. A calculation of  the integral in \eq{eq:integral1expanded} with $\Delta_2=0$ is much simpler and leads to same result as in \eq{eq:i1answer}. Thus we get the final result
\begin{align}\label{2loopanswerAppendix}
\text{Im} \, \mathcal{M}^{(2)}_{\text{QCD}} = \frac{g^6}{256 \pi^5}\frac{1}{s} \int \, &d^2\vc{l}_{1\perp}\d^2\vc{l}_{2\perp}\frac{\theta\left(s-\left(\sqrt{\Delta_1}+\sqrt{\Delta_2}+\sqrt{\Delta_{12}}\right)^2\right)}
{(\Delta_1\Delta_2)^2}\left[\ln\frac{s}{\Delta_{12}}-2\right]\, .
\end{align}
Note that this result has been derived assuming that $t=-\vc{q}_{\perp}^2=0$. Also note that to the leading order in $\Delta/s$ the theta function can be set to $1$.
\subsection{Modes in the EFT}
The two-loop graph in which first momentum is collinear ($l_1$) and second one is Glauber gluon ($l_2$) is equal to:
\begin{eqnarray}\label{2loopGnbarAppendix}
\text{Im} \, \mathcal{M}^{(2)}_{nG} = \frac{g^6}{64 \pi^5} \int \, d^4l_1 d^4l_2 \, \frac{\delta_+[(-l_1^+)(\sqrt{s}-l_1^-)-\Delta_1] \delta_+[(l_1^+-l_2^+)l_1^--\Delta_{12}]  \delta_+[\sqrt{s}\,l_2^--\Delta_2] \,\theta(\sqrt{s}+l_2^+)}{(l_1^+l_1^--\Delta_1)(-\Delta_2) ((l_1^++q^+)\,l_1^--\Delta_{1q}) (-\Delta_{2q})}\, .
\end{eqnarray}
The expression above can be found from expanding the full QCD graph in the given momentum region. The integration over $l_1^+, l_2^+, l_2^-$ can be performed using the three delta functions. As a result we get
\begin{eqnarray}
\text{Im} \, \mathcal{M}^{(2)}_{nG} =\frac{1}{s}\int\d^2\vc{l}_{1\perp}\d^2\vc{l}_{2\perp}\frac{\theta\left(s-\left(\sqrt{\Delta_{12}}+\sqrt{\Delta_{1}}\right)^2\right)}
{\Delta_1\Delta_2\Delta_{1q}\Delta_{2q}}\,I_{l_1^-}\, ,
\end{eqnarray}
where
\begin{eqnarray}
I_{l_1^-}=\int_{{{\Delta_{12}}/{\sqrt{s}}}}^{\sqrt{s}}\frac{\d l_1^-}{l_1^-}\frac{\Delta_{1q}}{\Delta_{1q}+\left(\frac{\Delta_1}{\sqrt{s}-l_1^-}+q^+\right)\,l_1^-}
=\ln\frac{s}{\Delta_{12}}+\frac{1}{2}\ln\frac{\Delta_{1q}}{\Delta_{1}}-\frac{\arctan U}{U}\, .
\end{eqnarray}
In the equation above the quantity $U$ equals to
\begin{eqnarray}
U=\sqrt{\frac{4\Delta_1\Delta_{1q}}{(t+\Delta_{1}+\Delta_{1q})^2}-1}\, .
\end{eqnarray}
For all the values of $t, \Delta_{1}, \Delta_{1q}$ consistent with their values as a function of $\vc{l}_{1\perp}, \vc{l}_{2\perp}, m$ the quantity $U>0$. Thus, we get the following final result for this two-loop $nG$ loop integral:
\begin{align}\label{2loopGnbarAppendix}
\text{Im} \, \mathcal{M}^{(2)}_{nG} = \frac{g^6}{256 \pi^5} \frac{1}{s}\int \, &d^2\vc{l}_{1\perp}\d^2\vc{l}_{2\perp}\frac{\theta\left(s-(\sqrt{\Delta_{12}}+\sqrt{\Delta_{1}})^2\right)}{\Delta_1\Delta_2\Delta_{1q}\Delta_{2q}}\left[\ln\frac{s}{\Delta_{12}}+\frac{1}{2}\ln\frac{\Delta_{1q}}{\Delta_{1}}-\frac{\arctan U}{U}\right].
\end{align}
Similar calculations for the $\text{Im} \, \mathcal{M}^{(2)}_{G\bar{n}}$ graph leads to an identical result.
\section{Overlap subtraction formula}
\noindent Here, we derive a master formula to account for overlaps given a number of regions. We start by stating certain assumptions about the construction \cite{jantzen}. In the following, a given region is denoted by $R_i$ and the full domain of integration by $R$. We have
\begin{align}
R_i \cap... \cap R_j = \emptyset , \quad R_i \cup ... \cup R_j = R \quad
\end{align}
We also assume that expansions commute; for example, $\mathcal{M}^{(1)}_{n\bar{n}} = \mathcal{M}^{(1)}_{\bar{n}n}$ as easily found in our explicit calculations. The last important property is that an integral converges absolutely within any region if the integrand is expanded appropriately. First, the full integral for $N$-modes is identically equal to
\begin{align}
\int_R \, dl\,  I =  \sum_{i=1}^{N} \int_{R_i} \, dl \, I_i
\end{align}
With some work, the integral in any given region can be written identically as follows
\begin{align}
\nonumber
\int_{R_i} \, dl \, I_i &= \int_{R} \, dl \, I_i - \int_{R} \, dl\,  \sum_{j\neq i}^{N} \, I_{ji} + \int_{R} \, dl\, \sum_{j\neq i}^{N} \sum_{k\neq i}^{N} \, I_{jki} - \text{4-fold overlaps} + ... \\
&+ \int_{R_i} \, dl \, \sum_{j\neq i}^{N} I_{ji} - \int_{R_i} \, dl \, \sum_{j\neq i}^{N} \sum_{k\neq i}^{N} I_{jki} + \text{4-fold overlaps} - ...
\end{align}
Notice that the sums are contstrained to avoid double counting under the assumption of commuting expansions. For example, if we have a total of four modes
\begin{align}
\nonumber
\int_{R_1} \, dl \, I_1 &= \int_{R} \, dl \, I_1 - \int_{R} \, dl\,  \sum_{i\neq 1}^{4} \, I_{i1} + \int_{R} \, dl\, \sum_{i\neq 1}^{4} \sum_{j\neq 1}^{4} \, I_{ij1} - \int_{R} \, dl \, I_{3241} \\
&+ \int_{R_1} \, dl \, \sum_{i\neq 1}^{4} I_{i1} - \int_{R_1} \, dl \, \sum_{i\neq 1}^{4} \sum_{j\neq 1}^{4} I_{ij1} + \int_{R_1} \, dl \, I_{3241}
\end{align}
Now the full result is obtained by adding all the modes to yield
\begin{align}
\int_{R} \, dl \, I = \int_R \, dl \, \sum_{i=1}^{N} I_i - \int_R \, dl \, \sum_{i=1}^{N} \sum_{j>i}^{N} \, I_{ij} + \int_R \, dl \, \sum_{i=1}^{N} \sum_{j>i}^{N} \sum_{k>j}^{N} \, I_{ijk} - \text{four overlaps} + ...
\end{align}
Note that the order of subscript indices does not matter because expansions commute.


\end{document}